\documentclass[11pt, a4paper]{article}

\usepackage{amsmath}
\usepackage{amssymb}
\usepackage{amsbsy}

\usepackage{geometry}
\usepackage{color}
\usepackage{graphicx}
\usepackage{cite}

\newcommand{\abs}[1]{\left\lvert #1\right\rvert}
\newcommand{\veps}{\varepsilon}

\newcommand{\bH}{\mathbf{H}}
\newcommand{\bG}{\mathbf{G}}
\newcommand{\bo}{\boldsymbol{0}}
\newcommand{\bp}{\mathbf{p}}
\newcommand{\bq}{\mathbf{q}}

\newcommand{\whPsi}{\widehat\Psi}
\newcommand{\whp}{\widehat{p}}
\newcommand{\whq}{\widehat{q}}
\newcommand{\whbp}{\widehat{\bp}}
\newcommand{\whbq}{\widehat{\bq}}

\newcommand{\whc}{\widehat{c}}

\newcommand{\wtphi}{\widetilde{\phi}}
\newcommand{\wtc}{\widetilde{c}}
\newcommand{\wtPsi}{\widetilde{\Psi}}
\newcommand{\wtT}{\widetilde{T}}
\newcommand{\wtU}{\widetilde{U}}
\newcommand{\wtQ}{\widetilde{Q}}
\newcommand{\wtp}{\widetilde{p}}
\newcommand{\wtq}{\widetilde{q}}
\newcommand{\wtbp}{\widetilde{\bp}}
\newcommand{\wtbq}{\widetilde{\bq}}
\newcommand{\wtG}{\widetilde{\bG}}
\newcommand{\wtH}{\widetilde{\bH}}
\newcommand{\wtF}{\widetilde{F}}
\newcommand{\wtFf}{\widetilde{\Ff}}
\newcommand{\wtC}{\widetilde{C}}
\newcommand{\wtM}{\widetilde{M}}

\newcommand{\bbR}{\mathbb{R}}

\newcommand{\Ff}{\mathcal{F}}
\newcommand{\Rr}{\mathcal{R}}
\newcommand{\Aa}{\mathcal{A}}
\newcommand{\Bb}{\mathcal{B}}

\newcommand{\fp}{\mathfrak{p}}
\newcommand{\fq}{\mathfrak{q}}

\newcommand{\diag}{\mathrm{diag}}
\newcommand{\Tr}{\mathrm{Tr}}
\newcommand{\re}{\mathrm{Re}}

\DeclareMathOperator{\sn}{\mathrm{sn}}
\DeclareMathOperator{\cn}{\mathrm{cn}}
\DeclareMathOperator{\dn}{\mathrm{dn}}

\allowdisplaybreaks

\begin{document}

\title{The vector form of Kundu-Eckhaus  equation and its simplest solutions}

\author{Aleksandr O. Smirnov, Aleksandra A. Caplieva\\[5pt]
{\sl St.-Petersburg State University of Aerospace Instrumentation}\\
{\sl B. Morskaya str., 67A,  St.-Petersburg, 1900000, Russia}
}






\maketitle

\begin{abstract}
In our work a hierarchy of integrable vector nonlinear differential equations depending on the functional parameter $r$ is constructed using a monodromy matrix. The first equation of this hierarchy for $r=\alpha(\bp^t\bq)$ is  vector analogue of the Kundu-Eckhaus equation. When $\alpha=0$, the equations of this hierarchy turn into equations of the Manakov system hierarchy. New elliptic solutions to vector analogue of the Kundu-Eckhaus and Manakov system are presented. In conclusion, it is shown that there exist  linear transformations of solutions to vector integrable nonlinear equations into other solutions to the same equations.
\end{abstract}

Keywords: Monodromy matrix, spectral curve, derivative nonlinear Schrodinger equation, vector integrable nonlinear equation

\section*{Introduction}

It is well known that the derived nonlinear Schrodinger equations \cite{KN, DNLS79, GI83, GI83a, Wad99, Sm21}  have numerous applications in various fields of physics and mathematics. In this regard, studies of various types of solutions to these equations are constantly being carried out (see, for example, \cite{KN18, KN19, DNLS20, Yan20, DNLS21, KN21}). At the same time, it should be noted that along with the Kaup-Newell \cite{KN} equation
\begin{equation} \label{eq.KN}
ip_z+p_{tt}+i(\abs{p}^2p)_t=0,
\end{equation}
Chen-Lee-Liu equation \cite{DNLS79}
\begin{equation} \label{eq.CLL}
ip_z+p_{tt}+i\abs{p}^2 p_t=0,
\end{equation}
and Gerdjikov-Ivanov equation \cite{GI83, GI83a}
\begin{equation} \label{eq.GI}
ip_z+p_{tt}-ip^2p_t^\ast+\dfrac12\abs{p}^4p=0,
\end{equation}
there exists the Kundu-Eckhaus equation \cite{DNLS84, CalE, Kundu, HeKundu15, HeKundu22}
\begin{equation} \label{eq.Kundu}
ip_z+p_{tt}-2\sigma\abs{p}^2p+\alpha^2\abs{p}^4p+2i\alpha\partial_t \left(\abs{p}^2\right) p=0, \quad \sigma=\pm1.
\end{equation}
Equation \eqref{eq.Kundu}, as well as  equations \eqref{eq.KN}-\eqref{eq.GI}, contains  first derivatives and also has numerous applications.

However, there is a significant difference between  equations \eqref{eq.KN}-\eqref{eq.GI} and \eqref{eq.Kundu}. The first three equations are consequences of compatibility conditions  of Lax pairs with quadratic in spectral parameter Lax operators. Equation \eqref{eq.Kundu}, in contrast to  equations \eqref{eq.KN}-\eqref{eq.GI}, is a result of the gauge transformation
\begin{equation*}
p=\whp e^{-i\theta},\quad \theta=\alpha\int \abs{p}^2 dt
\end{equation*}
of a solution $\whp$ to the nonlinear Schr{\"o}dinger equation
\begin{equation*}
i\whp_z+\whp_{tt}-2\sigma\abs{\whp}^2\whp=0.
\end{equation*}

The ever-growing traffic in  networks requires finding ways to increase  bandwidth of optical fibers.
Therefore, researchers are actively working on vector models of nonlinear optical waves propagation  \cite{Vnls17, Vnls18, Vnls20}. Many of these models have been known for a long time. One such model is the Manakov system \cite{Man74, EEK00, EEI07, WE07, GSM20epj, Sm20epj}
\begin{equation} \label{eq.Man}
\begin{aligned}
&\partial_zp_1=i\partial^2_tp_1-2i\sigma(\abs{p_1}^2+\abs{p_2}^2)p_1,\\
&\partial_zp_2=i\partial^2_tp_2-2i\sigma(\abs{p_1}^2+\abs{p_2}^2)p_2.
\end{aligned}
\end{equation}

Let us note that vector nonlinear Schr{\"o}dinger equations also have derivative forms, one of which is the equations \cite{Vdnls79, Vdnls09, Vdnls10, Vdnls16, Vdnls19}
\begin{equation}
\begin{aligned}
&i\partial_zp_1=-\partial^2_{t}p_1-\dfrac{2i}3\partial_t\left[\left(\abs{p_1}^2+\abs{p_2}^2\right)p_1\right],\\
&i\partial_zp_2=-\partial^2_{t}p_2-\dfrac{2i}3\partial_t\left[\left(\abs{p_1}^2+\abs{p_2}^2\right)p_2\right].
\end{aligned}
\end{equation}

In contrast to the above works, we decided to investigate a vector analogue of the Kundu-Eckhaus equation.
In this paper, we use a monodromy matrix to derive  equations from vector analogue of the Kundu-Eckhaus equation hierarchy and construct the simplest nontrivial solutions to first equation. We hope that the vector equation we have obtained, as well as the scalar one, will have numerous applications in physics and mathematics.

The work consists of an introduction, four sections, and concluding remarks. In the first section, we define the Lax operator
\begin{equation*}
i\Psi_{t}+U\Psi=\bo, 
\end{equation*}
which depends on a functional parameter $r\in\bbR$, and investigate  properties of  corresponding monodromy matrix. Since the spectral curve equation  is a characteristic equation of the monodromy matrix \cite{Dub85e}, it is not difficult to obtain  properties of the spectral curves equations from  properties of the monodromy matrix. As in the case of the Manakov system \cite{Sm20epj}, the spectral curves equations are quite cumbersome, so we will not give them in this paper. Let us note only that, as is shown in \cite{Sm20epj},  linear dependence of the functions $p_j$ leads to  factorization of the spectral curve equation  into separate components. Therefore, from our point of view, solutions with linearly independent $p_j$ are more  interesting to study and use in applications.

In section 2 we derive stationary equations for multiphase solutions. These equations are analogs of the Novikov equations for the Korteweg-de Vries hierarchy. Also in this section, we define a hierarchy of the second operators of the Lax pair
\begin{equation*}
i\Psi_{z_k}+W_k\Psi=\bo,
\end{equation*}
which depend on the functional parameter  $r_k$: $\partial_t r_k=\partial_{z_k}r$. The Lax pair compatibility conditions give a hierarchy of vector derivative  nonlinear Schr{\"o}dinger equations with an additional functional parameter $\phi$: $r=\partial_t\phi$ and $r_k=\partial_{z_k}\phi$. When $r=\alpha(\bp^t\bq)$ these equations are vector analogue of the Kundu-Eckhaus equation and its higher forms. Another choice of the functional parameter leads to other vector nonlinear equations. For $\phi\equiv0$ these equations will turn into equations from the Manakov hierarchy \cite{Sm20epj}. Let us note that an existence of a Lax pair makes it possible to use the Darboux transformation to construct new solutions to vector analogue of the Kundu-Eckhaus equation.

In section 3, we construct one-phase solutions to vector analogue of the Kundu-Eckhaus equation.
The first three solutions are expressed in terms of elliptic Jacobi functions, and for $\phi\equiv0$ they will be new elliptic solutions to the Manakov system.
Let us recall, that obtained in \cite{Sm20epj} elliptic solutions to the Manakov system  were expressed in terms of Weierstrass functions. 
Then we construct solutions expressed in terms of hyperbolic functions.  At the end of the section, we consider one-phase two-gap solutions. Despite the fact that in the last case the spectral curve has a genus equal to 2, the corresponding solution is a traveling wave.

In the section 4, we show that there exist linear transformations of solutions to vector integrable nonlinear equations into other solutions to the same equations. Original and transformed solutions are associated with the same spectral curve, but they correspond to different Baker-Akhiezer functions. One of these Baker-Achiezer functions differs from the other by an orthogonal matrix factor. I.e., the Baker-Achiezer function for considered vector nonlinear Schr{\"o}dinger equation is determined up to orthogonal transformation.

\section{The monodromy matrix and its properties}

Let  first equation of a Lax pair have the form
\begin{equation} \label{lax.1}
i\Psi_{t}+U\Psi=\bo, 
\end{equation}
where
\begin{equation*}
U=U_0+rJ,\quad U_0=-\lambda J+Q,
\end{equation*}
\begin{equation*}
J=\dfrac13\begin{pmatrix} 2 & \bo^t \\ \bo &-I \end{pmatrix},\quad
Q=\begin{pmatrix} 0 & \bp^t \\ -\bq & \bo \end{pmatrix},
\end{equation*}
$\bp^t=(p_1,p_2)$, $\bq^t=(q_1,q_2)$, $I$ is  identity matrix, $r\in\bbR$ is a some function, and $\lambda$ is a spectral parameter.

Following \cite{Dub85e, Sm20epj}, assume that there exists a monodromy matrix $M$ such that the matrix function $\whPsi=M\Psi$ is also an eigenfunction of  Lax operator \eqref{lax.1}. Then the matrix $M$ satisfies the equation
\begin{equation} \label{eq.M}
iM_{t}+UM-MU=\bo.
\end{equation}
In the case of a finite-gap matrix potential $Q$, the monodromy matrix $M$ is a polynomial in the spectral parameter $\lambda$ \cite{Dub85e, Sm20epj}
\begin{equation} \label{def.M}
M=\sum_{j=0}^nm_j(t)\lambda^j.
\end{equation}

Substituting \eqref{def.M} in \eqref{eq.M} and simplifying, we get that the matrix $M$ has the following structure
\begin{equation*}
M=V_n+\sum_{k=1}^{n-1}c_kV_{n-k}+c_nU_0+J_n,
\end{equation*}
where $V_1=\lambda U_0 +V_1^0$, 
\begin{gather*}
V_{k+1}=\lambda V_k+V_{k+1}^0,\quad
V_k^0=\begin{pmatrix}-\Ff _k& \bH_k^t\\
\bG_k & F_k
\end{pmatrix}, \quad k\ge1,\\
J_{n}=\begin{pmatrix}
-c_{n+1}-c_{n+2} & 0 &0\\
0& c_{n+1} & c_{n+3}\\
0& c_{n+4} & c_{n+2}
\end{pmatrix},
\end{gather*}
$c_j$ are some constants, $\Ff_k=\Tr F_k$.

From  equation \eqref{eq.M} it follows that  elements of the matrices $V_k^0$ satisfy  recurrence relations
\begin{equation} \label{eq.rec}
\begin{aligned}
&\bH_1=i\partial_t\bp+r\bp,\\
&\bG_1=i\partial_t \bq-r\bq,\\
& F_k=-i\partial_t^{-1}\left(\bG_k\bp^t+\bq\bH_k^t\right),\\
&\bH_{k+1}=i\partial_t\bH_k+r\bH_k+\left(F_k^t+\Ff_k I\right)\bp,\\
&\bG_{k+1}=-i\partial_t\bG_k+r\bG_k-\left(F_k+\Ff_k I\right)\bq.
\end{aligned}
\end{equation}

In particular, $F_1=\bq\bp^t$, $\Ff_1=\bp^t\bq=\bq^t\bp$,
\begin{align*}
&\bH_2=-\partial_t^2 \bp+2ir\partial_t\bp+\left(2\bp^t\bq+r^2+i\partial_t r\right)\bp,\\
&\bG_2=\partial_t^2\bq+2ir\partial_t \bq-\left(2\bp^t\bq+r^2-i\partial_t r\right)\bq,\\
&F_2=2\left(\bq\bp^t\right)r+i\left(\bq\partial_t\bp^t-\partial_t\bq\bp^t\right)=\bq\bH_1^t-\bG_1\bp^t,\\
&\Ff_2=2(\bp^t\bq)r+i(\bq^t\partial_t\bp-\bp^t\partial_t\bq)=\bH_1^t\bq-\bp^t\bG_1,\\
&\bH_3=-i\partial_t^3\bp-3r\partial_t^2\bp+3i\left(\bp^t\bq+r^2+i\partial_t r\right)\partial_t\bp\\
&\qquad+\left(3i \partial_t\bp^t\bq+r^3+6r\bp^t\bq +3ir\partial_t r-\partial_t^2 r\right)\bp,\\
&\bG_3=-i\partial_t^3\bq+3r\partial_t^2\bq+3i\left(\bp^t\bq+r^2-i\partial_t r\right)\partial_t\bq\\
&\qquad+\left(3i\bp^t\partial_t\bq-r^3-6r\bp^t\bq+3ir\partial_r r+\partial_t^2 r\right)\bq,\\
&F_3=3(\bq\bp^t)r^2+3i(\bq\partial_t\bp^t-\partial_t\bq\bp^t)r+\partial_t\bq\partial_t\bp^t\\
&\qquad-\bq\partial_t^2\bp^t-\partial_t^2\bq\bp^t+3(\bp^t\bq)\bq\bp^t\\
&\qquad= \bq\bH_2^t-\bG_2\bp^t-\bG_1\bH_1^t-(\bp^t\bq)\bq\bp^t,\\
&\Ff_3=3(\bp^t\bq)r^2+3i(\bq^t\partial_t\bp-\bp^t\partial_t\bq)r+3(\bp^t\bq)^2\\
&\qquad+\partial_t\bp^t\partial_t\bq-\bp^t\partial_t^2\bq-\bq^t\partial_t^2\bp.
\end{align*}

For $r\equiv0$, all the above equations turn into  corresponding equations for the Manakov system \cite{Sm20epj}.

\section{Stationary and evolutionary equations} \label{sec.equations}

From equation \eqref{eq.M} the following stationary equations follow:
\begin{equation} \label{eq.stat}
\begin{aligned}
&\bH_{n+1}+\sum_{k=1}^{n}c_k\bH_{n+1-k}+C_n^t\bp=\bo,\\
&\bG_{n+1}+\sum_{k=1}^{n}c_k\bG_{n+1-k}-C_n\bq=\bo,
\end{aligned}
\end{equation}
where
\begin{equation*}
C_n=\begin{pmatrix} 2c_{n+1}+c_{n+2} & c_{n+3} \\ c_{n+4} & c_{n+1}+2c_{n+2} \end{pmatrix}.
\end{equation*}
All multiphase solutions to evolutionary integrable nonlinear equations are simultaneously solutions to some stationary equations.

In the case of reduction $\bq=\sigma\bp^\ast$ ($\sigma=\pm1$), the following identities 
\begin{gather*}
\bG_k^\ast=-\sigma \bH_k,\quad \bH_k^\ast=-\sigma \bG_k,\quad F_k^\ast=F_k^t,\quad \Ff_k\in\bbR
\end{gather*}
follow from the recurrence relations \eqref{eq.rec}.
Therefore, the constants $c_j$  in stationary equations \eqref{eq.stat} must satisfy the following conditions: $c_k\in\bbR$ ($1\leq k\leq n+2$), $c_{n+4}=c_{n+3}^\ast$.

Let a second operator of a Lax pair have the form
\begin{equation} \label{lax.2}
i\Psi_{z_k}+W_k\Psi=\bo
\end{equation}
where $W_k=V_k+r_kJ$.
Then from compatibility condition  of  equations \eqref{lax.1} and \eqref{lax.2}, or from equation
\begin{equation*}
i\partial_tW_k-i\partial_{z_k}U+UW_k-W_kU=\bo
\end{equation*}
the evolutionary integrable nonlinear equations
\begin{equation}
i\partial_{z_k}\bp=\bH_{k+1}-r_k\bp,\quad
i\partial_{z_k}\bq=\bG_{k+1}+r_k\bq, \label{eq.dzk}
\end{equation}
and the additional relation
\begin{equation}
\partial_{z_k}r=\partial_t r_k \label{eq.rtz}
\end{equation}
follow.
It follows from \eqref{eq.rtz} that there exists a function $\phi$ such that
\begin{equation*}
r=\partial_t\phi,\quad r_k=\partial_{z_k}\phi.
\end{equation*}

The first systems of integrable nonlinear equations from  hierarchy \eqref{eq.dzk} have the form
\begin{equation} \label{eq.man}
\begin{aligned}
&i\partial_{z_1}\bp=-\partial_t^2 \bp+2ir\partial_t\bp+\left(2\bp^t\bq+r^2+i\partial_t r-r_1\right)\bp,\\
&i\partial_{z_1}\bq=\partial_t^2\bq+2ir\partial_t \bq-\left(2\bp^t\bq+r^2-i\partial_t r-r_1\right)\bq,
\end{aligned}
\end{equation}
and
\begin{equation} \label{eq.mkdv}
\begin{aligned}
&\partial_{z_2}\bp=-\partial_t^3\bp+3ir\partial_t^2\bp+3\left(\bp^t\bq+r^2+i\partial_t r\right)\partial_t\bp\\
&\qquad+\left(3 \partial_t\bp^t\bq-ir^3-6ir\bp^t\bq +3r\partial_t r+i\partial_t^2 r+ir_2\right)\bp,\\
&\partial_{z_2}\bq=-\partial_t^3\bq-3ir\partial_t^2\bq+3\left(\bp^t\bq+r^2-i\partial_t r\right)\partial_t\bq\\
&\qquad+\left(3\bp^t\partial_t\bq+ir^3+6ir\bp^t\bq+3r\partial_r r-i\partial_t^2 r-ir_2\right)\bq.
\end{aligned}
\end{equation}
Equation \eqref{eq.man} is one of a vector derivative nonlinear Schr{\"o}dinger equations, and \eqref{eq.mkdv} is a  vector modified Korteweg-de Vries equation. Both equations are parametrized by an arbitrary real function $\phi$.

From \eqref{eq.rec} and \eqref{eq.dzk} the following equalities
\begin{equation*}
\partial_{z_k} F_1=\partial_t F_{k+1}\quad\text{and}\quad \partial_{z_k} \Ff_1=\partial_t \Ff_{k+1}
\end{equation*}
follow. 
 Therefore, there exist functions $\Phi$ and $\wtphi$ such that
\begin{equation*}
F_1=\partial_t\Phi,\quad F_{k+1}=\partial_{z_k}\Phi \quad\text{and}\quad \Ff_1=\partial_t \wtphi,\quad \Ff_{k+1}=\partial_{z_k}\wtphi.
\end{equation*}
Therefore, if we put  $\phi=\alpha\wtphi$ or 
\begin{equation} \label{eq.r.F}
r=\alpha\Ff_1 \quad\text{and}\quad r_k=\alpha\Ff_{k+1}, 
\end{equation}
then  equations \eqref{eq.dzk}, \eqref{eq.r.F} will determine  evolutionary integrable nonlinear equations from the hierarchy of a vector analogue of the Kundu-Eckhaus equation.

In particular, for $k=1$, or for
\begin{equation*}
r=\alpha\Ff_1=\alpha(\bp^t\bq)\quad\text{and}\quad  r_1=\alpha\Ff_2=2\alpha^2(\bp^t\bq)^2+i\alpha(\bq^t\partial_t\bp-\bp^t\partial_t\bq),
\end{equation*}
equation \eqref{eq.man} will have the form
\begin{equation} \label{eq.man.new}
\begin{aligned}
&i\partial_{z_1}\bp=-\partial_t^2 \bp+2i\alpha(\bp^t\bq)\partial_t\bp+\left(2\bp^t\bq-\alpha^2(\bp^t\bq)^2+2i\alpha\bp^t\partial_t\bq\right)\bp,\\
&i\partial_{z_1}\bq=\partial_t^2\bq+2i\alpha(\bp^t\bq)\partial_t \bq-\left(2\bp^t\bq-\alpha^2(\bp^t\bq)^2-2i\alpha\partial_t\bp^t\bq\right)\bq.
\end{aligned}
\end{equation}
When $\bq=S \bp^\ast$,  $S=\diag(\sigma_1,\sigma_2)$, $\sigma_j=\pm1$, equations \eqref{eq.man.new} transform into a vector analogue of the Kundu-Eckhaus equation.
It is not difficult to see that for $\alpha=0$  equations \eqref{eq.man.new} transform into the Manakov system \cite{Sm20epj}.

\section{One-phase solutions} \label{sec.solutions}

\subsection{Solutions in elliptic Jacobi functions}

For $n=1$   stationary equations have the form
\begin{equation} \label{n1.stat.GH}
\begin{aligned}
&\bH_2+c_1\bH_1+C_1^t\bp=\bo,\\
&\bG_2+c_1\bG_1-C_1\bq=\bo
\end{aligned}
\end{equation}
or (for $c_4=c_5=0$ and $r=\alpha\Ff_1$, $r_1=\alpha\Ff_2$)
\begin{equation} \label{n1.stat}
\begin{aligned}
&\partial_t^2p_1=i(c_1+2\alpha(\bp^t\bq))\partial_t p_1\\
&\qquad+(2c_2+c_3+(2+c_1\alpha)\bp^t\bq+\alpha^2(\bp^t\bq)^2+i\alpha\partial_t(\bp^t\bq))p_1,\\
&\partial_t^2p_2=i(c_1+2\alpha(\bp^t\bq))\partial_t p_2\\
&\qquad+(c_2+2c_3+(2+c_1\alpha)\bp^t\bq+\alpha^2(\bp^t\bq)^2+i\alpha\partial_t(\bp^t\bq))p_2,\\
&\partial_t^2q_1=-i(c_1+2\alpha(\bp^t\bq))\partial_t q_1\\
&\qquad+(2c_2+c_3+(2+c_1\alpha)\bp^t\bq+\alpha^2(\bp^t\bq)^2-i\alpha\partial_t(\bp^t\bq))q_1,\\
&\partial_t^2q_2=-i(c_1+2\alpha(\bp^t\bq))\partial_t q_2\\
&\qquad+(c_2+2c_3+(2+c_1\alpha)\bp^t\bq+\alpha^2(\bp^t\bq)^2-i\alpha\partial_t(\bp^t\bq))q_2.
\end{aligned}
\end{equation}
Replacing  functions $p_j$  and $q_j$  by formulas
\begin{equation*}
p_j=\whp_j e^{i\theta},\quad q_j=\whq_j e^{-i\theta},\quad \partial_t\theta=\dfrac12c_1+\alpha\bp^t\bq,
\end{equation*}
we obtain the following equalities
\begin{equation} \label{wwpq.tt}
\begin{aligned}
&\partial_t^2\whp_1=\left(2\whbp^t\whbq+2c_2+c_3-\dfrac14c_1^2\right)\whbp_1,\\
&\partial_t^2\whp_2=\left(2\whbp^t\whbq+c_2+2c_3-\dfrac14c_1^2\right)\whbp_2,\\
&\partial_t^2\whq_1=\left(2\whbp^t\whbq+2c_2+c_3-\dfrac14c_1^2\right)\whbq_1,\\
&\partial_t^2\whq_2=\left(2\whbp^t\whbq+c_2+2c_3-\dfrac14c_1^2\right)\whbq_2.
\end{aligned}
\end{equation}
It is not difficult to see that  functions $\whp_j$ and $\whq_j$ are solutions of the same second-order linear differential equations. Hence, their products $u_j=\whp_j\whq_j$ satisfy the corresponding Appel's equations (\cite{WWe}, Part II, Chapter 14, Example 10; \cite{AP})
\begin{equation} \label{eq.u12}
\begin{aligned}
&\partial_t^3u_1-(8u_1+8u_2+8c_2+4c_3-c_1^2)\partial_t u_1 -4\partial_t(u_1+u_2)u_1=0,\\
&\partial_t^3u_2-(8u_1+8u_2+4c_2+8c_3-c_1^2)\partial_t u_2 -4\partial_t(u_1+u_2)u_2=0.
\end{aligned}
\end{equation}
Let us introduce the notations: $u_1+u_2=u$, $u_1-u_2=v$. In these notations,  equations \eqref{eq.u12} take the form
\begin{equation} \label{eq.uv.t}
\begin{aligned}
&\partial_t^3u+(c_1^2-6c_2-6c_3-12u)\partial_tu=2(c_2-c_3)\partial_t v,\\
&\partial_t^3v+(c_1^2-6c_2-6c_3-8u)\partial_tv=2(c_2-c_3+2v)\partial_t u.
\end{aligned}
\end{equation}
The simplest solutions of  equations \eqref{eq.uv.t} can be obtained when $v=(c_3-c_2)/2$. In this case, the function $u$ satisfies the equation
\begin{equation*}
\partial_t^3u+(c_1^2-6c_2-6c_3-12u)\partial_tu=0
\end{equation*}
or
\begin{equation} \label{u.dt2}
\partial_t^2u+(c_1^2-6c_2-6c_3)u-6u^2=\whc_1,
\end{equation}
where $\whc_1$ is an integration constant. Simplifying  relation \eqref{u.dt2}, we obtain the equation
\begin{equation} \label{eq.ut.n1}
(\partial_t u)^2=4u^3-(c_1^2-6c_2-6c_3)u^2+2\whc_1u+\whc_2,
\end{equation}
where $\whc_2$ is a second integration constant.
It is well known that solutions to  equation \eqref{eq.ut.n1} are elliptic functions or their degenerations.

It is easy to verify that one of the non-degenerate solutions to the equation \eqref{eq.ut.n1} has the form
\begin{equation} \label{u.sn}
u=k^2\sn^2(t;k)+\dfrac1{12}c_1^2-\dfrac12(c_2+c_3)-\dfrac13(1+k^2),
\end{equation}
where $\sn(t;k)$ is an elliptic Jacobi function \cite{Akhe, SFe}, that satisies the equation
\begin{equation*}
[\sn'(t)]^2=(1-\sn^2(t))(1-k^2\sn^2(t)).
\end{equation*}
The integration constant for solution \eqref{u.sn} is equal to
\begin{align*}
&\whc_1=\dfrac1{24}c_1^4-\dfrac12(c_2+c_3)c_1^2+\dfrac32(c_2+c_3)^2-\dfrac23(1-k^2+k^4),\\
&\whc_2=-\dfrac1{432}c_1^6+\dfrac1{24}(c_2+c_3)c_1^4-\dfrac14(c_2+c_3)^2c_1^2+\dfrac19(1-k^2+k^4)c_1^2\\
&\qquad+\dfrac12(c_2+c_3)^3-\dfrac23(c_2+c_3)(1-k^2+k^4)-\dfrac4{27}(2-3k^2-3k^4+2k^6).
\end{align*}

Knowing  functions $u$ and $v$, we obtain  functions $u_j$
\begin{equation} \label{uj.sn}
\begin{aligned}
&u_1=\dfrac12(u+v)=\dfrac{k^2}2\sn^2(t;k)+\dfrac{c_1^2}{24}-\dfrac{c_2}2-\dfrac{1+k^2}6,\\
&u_2=\dfrac12(u-v)=\dfrac{k^2}2\sn^2(t;k)+\dfrac{c_1^2}{24}-\dfrac{c_3}2-\dfrac{1+k^2}6.
\end{aligned}
\end{equation}

Thus,  functions $\whp_j$ and $\whq_j$ are solutions to the equations
\begin{equation} \label{eq.whp.sn}
\begin{aligned}
&\partial_t^2\whp_1=\left(2k^2\sn^2(t;k)-\dfrac23(1+k^2)-\dfrac1{12}c_1^2+c_2\right)\whp_1,\\
&\partial_t^2\whp_2=\left(2k^2\sn^2(t;k)-\dfrac23(1+k^2)-\dfrac1{12}c_1^2+c_3\right)\whp_2.
\end{aligned}
\end{equation}
Since  functions $\whq_j$ satisfy the same equations as $\whp_j$, and the Wronskian of these solutions are constant:
\begin{equation*}
W[\whp_j,\whq_j]=2iW_j,
\end{equation*}
 functions $\whp_j$ and $\whq_j$ are equal to
\begin{equation} \label{whpq.u}
\whp_j=\sqrt{u_j}\exp\left\{-iW_j\int\dfrac{dt}{u_j}\right\},\quad \whq_j=\sqrt{u_j}\exp\left\{iW_j\int\dfrac{dt}{u_j}\right\},
\end{equation}
where  $u_j$ are defined by formulas \eqref{uj.sn}. Substituting expressions \eqref{whpq.u} into  equation \eqref{eq.whp.sn} and simplifying, we get
\begin{equation} \label{eq.Wj}
W_j^2=\dfrac1{6912}(c_1^2-12c_{j+1}-4-4k^2)(c_1^2-12c_{j+1}+8-4k^2)(c_1^2-12c_{j+1}-4+8k^2).
\end{equation}

\begin{figure}[htbp]
\begin{center}
\begin{tabular}{p{0.45\textwidth}p{0.45\textwidth}}
\includegraphics[width=0.4\textwidth]{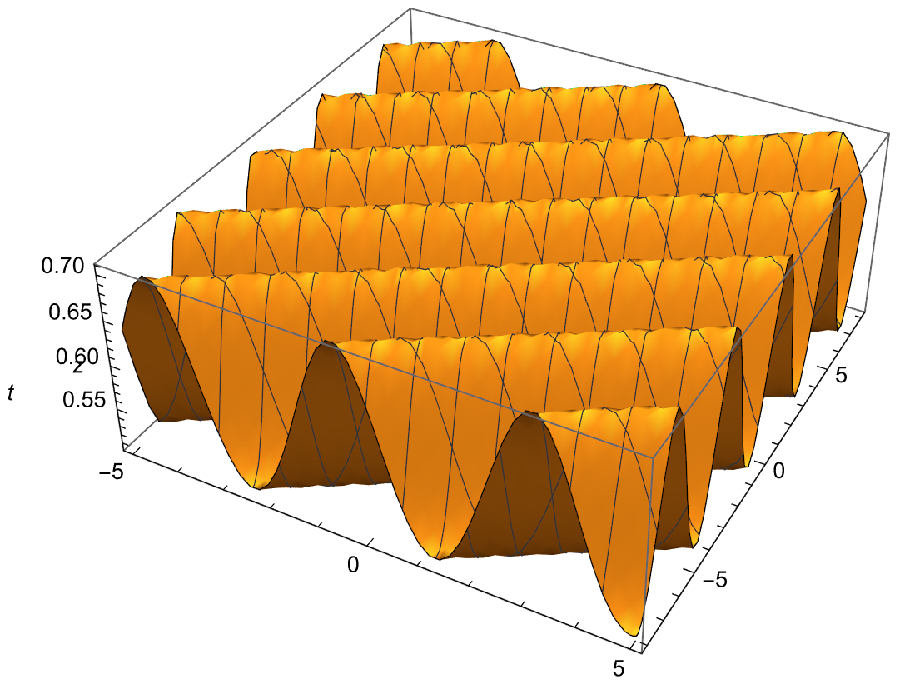}&
\includegraphics[width=0.4\textwidth]{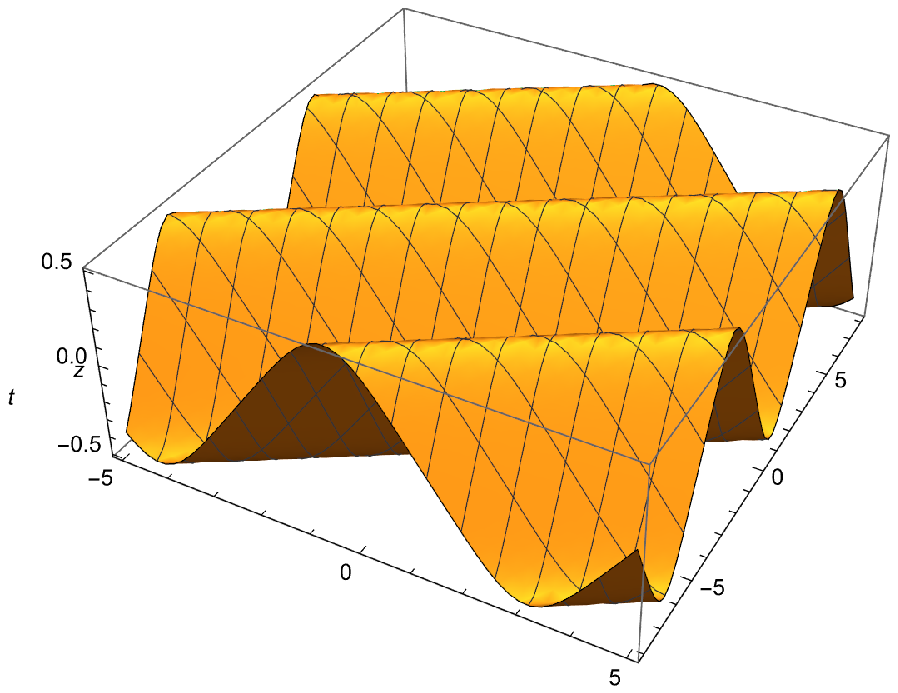}\\
\centering Magnitude $\fp_1(t-c_1z_1)$ & \centering Magnitude $\fp_2(t-c_1z_1)$
\end{tabular}
\caption{Magnitudes of  solutions  \eqref{sol.1} for $k=0.7$, $c_1=1$} \label{figsol.1}
\end{center}
\end{figure}

It follows from  equation \eqref{eq.Wj} that there are three cases when $\whp_j=\whq_j$ and $\whp_1\ne\whp_2$.
If
\begin{equation*}
c_2=\dfrac1{12}(c_1^2+8-4k^2),\quad c_3=\dfrac1{12}(c_1^2+8k^2-4),
\end{equation*}
then
\begin{equation*}
\whp_1=\whq_1=\dfrac{i}{\sqrt{2}}\dn(t;k),\quad \whp_2=\whq_2=\dfrac{ik}{\sqrt{2}}\cn(t;k).
\end{equation*}
In this case, the solution to  equations \eqref{eq.man.new} has the form
\begin{equation} \label{sol.1}
\begin{aligned}
&p_1=i\fp_1(t-c_1z_1)e^{i\theta},\quad q_1=-p_1^\ast,\\
&p_2=i\fp_2(t-c_1z_1)e^{i\theta},\quad q_2=-p_2^\ast,
\end{aligned}
\end{equation}
where
\begin{align*}
&\fp_1(T)=\dfrac{1}{\sqrt{2}}\dn(T;k), \quad \fp_2(T)=\dfrac{k}{\sqrt{2}}\cn(T;k),\\
&\theta=\dfrac{c_1}2t+\left(1-\dfrac{c_1^2}4\right) z_1+\alpha\int\left(k^2\sn^2(t-c_1 z_1;k)-\dfrac{1+k^2}2\right)dt.
\end{align*}
Magnitudes of  solutions  \eqref{sol.1} are shown in Fig.~\ref{figsol.1}.

For
\begin{equation*}
c_2=\dfrac1{12}(c_1^2+8-4k^2),\quad c_3=\dfrac1{12}(c_1^2-4k^2-4),
\end{equation*}
we have
\begin{equation*}
\whp_1=\whq_1=\dfrac{i}{\sqrt{2}}\dn(t;k),\quad \whp_2=\whq_2=\dfrac{k}{\sqrt{2}}\sn(t;k).
\end{equation*}
The corresponding solution to  equations \eqref{eq.man.new} has the form
\begin{equation} \label{sol.2}
\begin{aligned}
&p_1=i\fp_1(t-c_1z_1)e^{i\theta},\quad q_1=-p_1^\ast,\\
&p_2=\fp_2(t-c_1z_1)e^{i\theta},\quad q_2=p_2^\ast,
\end{aligned}
\end{equation}
where
\begin{align*}
&\fp_1(T)=\dfrac{1}{\sqrt{2}}\dn(T;k), \quad \fp_2(T)=\dfrac{k}{\sqrt{2}}\sn(T;k),\\
&\theta=\dfrac{c_1}2t+\left(1-k^2-\dfrac{c_1^2}4\right) z_1+\alpha\int\left(k^2\sn^2(t-c_1 z_1;k)-\dfrac{1}2\right)dt.
\end{align*}

If
\begin{equation*}
c_2=\dfrac1{12}(c_1^2+8k^2-4),\quad c_3=\dfrac1{12}(c_1^2-4k^2-4),
\end{equation*}
we have
\begin{equation*}
\whp_1=\whq_1=\dfrac{ik}{\sqrt{2}}\cn(t;k),\quad \whp_2=\whq_2=\dfrac{k}{\sqrt{2}}\sn(t;k).
\end{equation*}
In this case, the solution to  equations \eqref{eq.man.new} has the form
\begin{equation} \label{sol.3}
\begin{aligned}
&p_1=i\fp_1(t-c_1z_1)e^{i\theta},\quad q_1=-p_1^\ast,\\
&p_2=\fp_2(t-c_1z_1)e^{i\theta},\quad q_2=p_2^\ast,\\
\end{aligned}
\end{equation}
where
\begin{align*}
&\fp_1(T)=\dfrac{k}{\sqrt{2}}\cn(T;k), \quad \fp_2(T)=\dfrac{k}{\sqrt{2}}\sn(T;k),\\
&\theta=\dfrac{c_1}2t+\left(k^2-1-\dfrac{c_1^2}4\right) z_1+\alpha\int\left(k^2\sn^2(t-c_1 z_1;k)-\dfrac{k^2}2\right)dt.
\end{align*}

Dependency of solutions \eqref{sol.1}-\eqref{sol.3} from the variable $z_1$ was found from  equations~\eqref{eq.man.new}.

\subsection{Solutions in hyperbolic functions}

The equation \eqref{eq.ut.n1} is well studied. In particular, it has the following solution 
\begin{equation*}
u=a^2\tanh^2(at)+\dfrac1{12}\left((c_1^2-6c_2-6c_3)-8a^2\right).
\end{equation*}
Integration constants  for a given function $u$ are equal to
\begin{align*}
&\whc_1=\dfrac1{24}\left((c_1^2-6c_2-6c_3)^2-16a^4\right),\\
&\whc_2=\dfrac1{432}\left(4a^2+(c_1^2-6c_2-6c_3)\right)^2\left(8a^2-(c_1^2-6c_2-6c_3)\right).
\end{align*}
In this case,  functions $\whp_j$ and $\whq_j$ satisfy the following equations:
\begin{equation} \label{whp.tt.th}
\partial_t^2\whp_j=\left(2a^2\tanh^2(at)-\dfrac{4a^2}3-\dfrac{c_1^2}{12}+c_{j+1}\right)\whp_j
\end{equation}
and
\begin{equation} \label{eq.uj.th}
u_j=\whp_j\whq_j=\dfrac{a^2}2\tanh^2(at)+\dfrac1{24}\left(c_1^2-8a^2-12c_{j+1}\right).
\end{equation}
Let us recall that  functions $\whq_j$ also satisfy  equations \eqref{whp.tt.th}.

\begin{figure}[hbtp]
\begin{center}
\begin{tabular}{p{0.45\textwidth}p{0.45\textwidth}}
\includegraphics[width=0.4\textwidth]{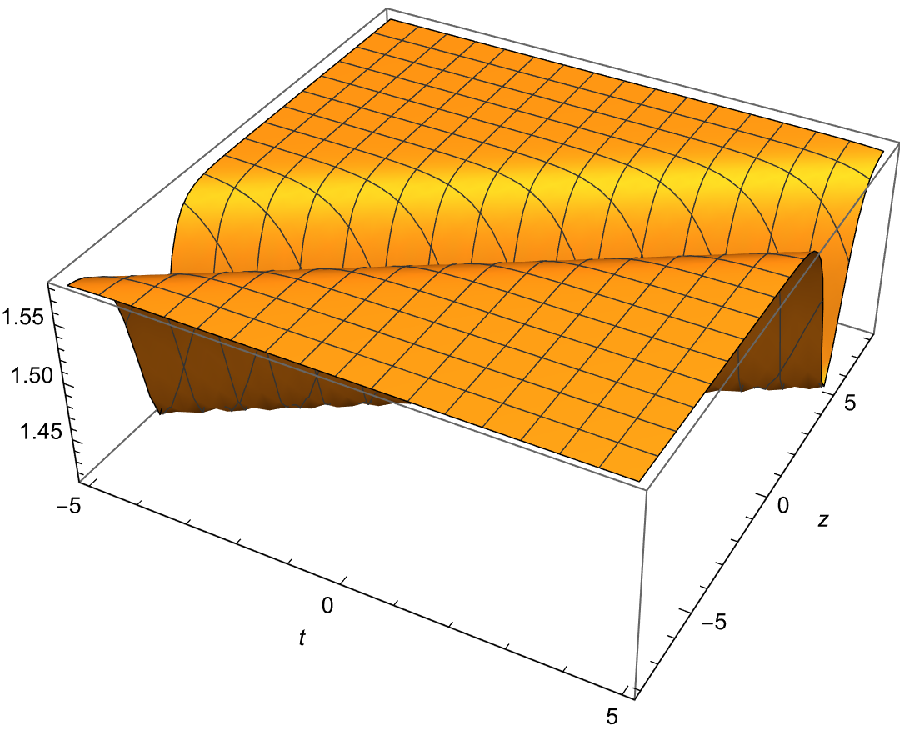}&
\includegraphics[width=0.4\textwidth]{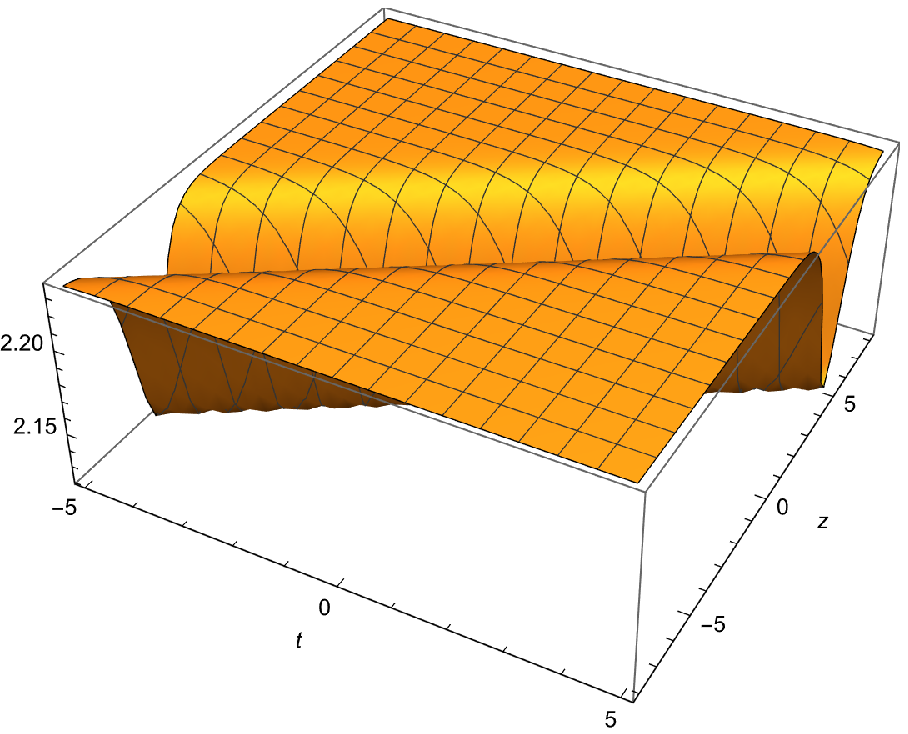}\\
\centering Magnitude $\abs{p_1(t,z_1)}$ & \centering Magnitude $\abs{p_2(t,z_1)} $
\end{tabular}
\caption{Magnitudes of  solutions  \eqref{sol.4} for $a=1$, $k_1=2$, $k_2=3$, $c_1=1$.} \label{figsol.2}
\end{center}
\end{figure}

Solving   equations \eqref{whp.tt.th} with conditions \eqref{eq.uj.th}, we get
\begin{equation}
\whp_j=\dfrac1{\sqrt2}(k_j+ia\tanh(at))e^{ik_j t},\quad
\whq_j=\dfrac1{\sqrt2}(k_j-ia\tanh(at))e^{-ik_j t},
\end{equation}
where
\begin{equation*}
k_j^2=\dfrac1{12}\left(c_1^2-8a^2-12c_{j+1}\right)\quad\text{or}\quad c_{j+1}=\dfrac1{12}\left(c_1^2-8a^2-12k_j^2\right).
\end{equation*}
The corresponding solution to  equations \eqref{eq.man.new} has the form
\begin{equation} \label{sol.4}
\begin{aligned}
&p_j=\dfrac1{\sqrt2}\left(k_j+ia\tanh[a(t-c_1z_1)]\right)e^{i\theta_j},\quad q_j=p_j^\ast,\\
&\theta_j=\left(\dfrac{c_1}2+k_j\right)t+m_j z_1+\alpha\int\left(a\tanh^2(at-ac_1z_1)+\dfrac{k_1^2+k_2^2}{2a}\right)dt,
\end{aligned}
\end{equation}
where
\begin{equation*}
m_j=-2a^2-\dfrac14(c_1+2k_j)^2-k_1^2-k_2^2-\alpha(k_1+k_2)(a^2+k_1^2-k_1k_2+k_2^2).
\end{equation*}
Magnitudes of  solutions  \eqref{sol.4} are shown in Fig.~\ref{figsol.2}.

\subsection{Two-gap one-phase solutions}

If $n=1$, $v\ne const,$ and $c_3\ne c_2$, then from \eqref{eq.uv.t} we have
\begin{equation} \label{v.u}
v=\dfrac1{2(c_2-c_3)}\left(u_{tt}-6u^2+(c_1^2-6c_2-6c_3)u\right)+C_1,
\end{equation}
and
\begin{multline} \label{u.t4}
u_{tttt} +2(c_1^2-6c_2-6c_3-10u)u_{tt}-10\left( u_t\right)^2+40u^3-12(c_1^2-6c_2-6c_3)u^2\\
+\left(c_1^4-12c_1^2(c_2+c_3)+8\left(4c_2^2+10c_2c_3+4c_3^3-C_1c_2+C_1c_3\right)\right)u+C_2=0,
\end{multline}
where $C_1$ and $C_2$ are  integration constants.

We can rewrite equation \eqref{u.t4}  in the form
\begin{equation} \label{eq.stat.g2}
I_2+2A I_1+(A^2-4B^2+8C_1 B)u+C_2=0,
\end{equation}
where $A=c_1^2-6(c_3+c_2)$, $B=c_3-c_2$,
\begin{equation} \label{int.g2}
I_2=u_{tttt} -20uu_{tt}-10\left(u_t\right)^2+40u^3,\quad I_1=u_{tt}-6u^2.
\end{equation}

It follows from equations \eqref{eq.stat.g2} and \eqref{int.g2} that the function $2u(t)$ is a two-gap potential of the Schr{\"o}dinger operator \cite{Nov74e, DNe}
\begin{equation} \label{op.Sch}
\psi_{tt}-2u\psi=E\psi.
\end{equation}
It is well known that linear independent solutions to  equation \eqref{op.Sch} with two-gap potential $2u(t)$  can be written as
\begin{equation} \label{sol.Sch}
\psi_{1,2}=\veps_{1,2}\sqrt{\Psi}\exp\left\{\pm\nu(E)\int\dfrac{dt}{\Psi}\right\},
\end{equation}
where
\begin{equation*}
\Psi=E^2+(\gamma_1-u)E+\gamma_2-\gamma_1 u-\dfrac14I_1,
\end{equation*}
and
\begin{equation*}
\gamma_1=\dfrac{A}2,\quad \gamma_2=\dfrac1{16}\left(A^2-4B^2+8C_1 B\right).
\end{equation*}

Substituing \eqref{sol.Sch} in \eqref{op.Sch} and simplifying, we obtain an equation for spectral curve of two-gap potential $u(t)$:
\begin{equation} \label{eq.curve.g2}
\nu^2=E^5+2\gamma_1E^4+(\gamma_1^2+2\gamma_2)E^3+\left(2\gamma_1\gamma_2-\dfrac18C_2\right)E^2+C_3E+C_4,
\end{equation}
where
\begin{align*}
&C_3=\dfrac18u_tu_{ttt}-\dfrac1{16}u_{tt}^2+\dfrac14(\gamma_1-5u)u_t^2+\dfrac54u^4-\gamma_1u^3+\gamma_2u^2+\dfrac18C_2u+\gamma_2^2-\dfrac18\gamma_1C_2,\\
&C_4=\dfrac1{64}u_{ttt}^2+\dfrac18(\gamma_1-3u)u_tu_{ttt}-\dfrac18uu_{tt}^2+\dfrac1{32}\left(C_2+16\gamma_2 u-24\gamma_1 u^2+40u^3+2u_t^2\right)u_{tt}\\
&\qquad +\dfrac18\left(2\gamma_1^2-2\gamma_2-10\gamma_1 u+15u^2\right)u_t^2-3u^5+\dfrac72\gamma_1 u^4-(\gamma_1^2+2\gamma_2)u^3\\
&\qquad+\dfrac1{16}(16\gamma_1\gamma_2-3C_2)u^2+\dfrac18\gamma_1 C_2 u-\dfrac18\gamma_2C_2.
\end{align*}

From \eqref{v.u} we have
\begin{equation} \label{eq.uj.g2}
\begin{aligned}
&u_1=\dfrac12(u+v)=-\dfrac1{4B}I_1+\left(\dfrac12-\dfrac{A}{4B}\right)u+\dfrac12C_1,\\
&u_2=\dfrac12(u-v)=\dfrac1{4B}I_1+\left(\dfrac12+\dfrac{A}{4B}\right)u-\dfrac12C_1.
\end{aligned}
\end{equation}
Therefore,  functions $\whp_j$ and $\whq_j$ are solutions to  equations
\begin{equation} \label{eq.whp.g2}
\begin{aligned}
&\partial_t^2\whp_1-2u\whp_1=-\dfrac14\left(A+2B\right)\whp_1,\\
&\partial_t^2\whp_2-2u\whp_2=-\dfrac14\left(A-2B\right)\whp_2.
\end{aligned}
\end{equation}

It is following from \eqref{op.Sch}, \eqref{sol.Sch}, \eqref{eq.uj.g2}, and \eqref{eq.whp.g2} that 
\begin{align*}
&\whp_1=\left.\veps_{11}\sqrt{\Psi}\exp\left\{\nu(E)\int\dfrac{dt}{\Psi}\right\}\right|_{E=E_1},\\
&\whq_1=\left.\veps_{12}\sqrt{\Psi}\exp\left\{-\nu(E)\int\dfrac{dt}{\Psi}\right\}\right|_{E=E_1},\\
&\whp_2=\left.\veps_{21}\sqrt{\Psi}\exp\left\{\nu(E)\int\dfrac{dt}{\Psi}\right\}\right|_{E=E_2},\\
&\whq_2=\left.\veps_{22}\sqrt{\Psi}\exp\left\{-\nu(E)\int\dfrac{dt}{\Psi}\right\}\right|_{E=E_2},
\end{align*}
where 
\begin{equation*}
\veps_{11}\veps_{12}=1/B, \quad\veps_{21}\veps_{22}=-1/B, \quad E_1=-(A+2B)/4, \quad E_2=-(A-2B)/4.
\end{equation*}

It is not difficult to see that  functions $\whp_j$ and $\whq_j$ will be bounded when $E_j$ satisfies to   conditions $\re(\nu(E_j))=0$.

Two-soliton potential of  operator \eqref{op.Sch} has the form ($b>a>0$)
\begin{equation} \label{u.twosol}
u(t)=\dfrac{(a^2-b^2)(b^2-a^2+a^2\cosh(2bt)+b^2\cosh(2at)}{2(b\cosh(bt)\cosh(at)-a\sinh(bt)\sinh(at))^2}.
\end{equation}
Substituting \eqref{u.twosol} into \eqref{eq.stat.g2}, we get
\begin{equation*}
A=-2(a^2+b^2), \quad C_1=-\dfrac{(a^2-b^2)^2-B^2}{2B},\quad C_2=0.
\end{equation*}
The spectral curve \eqref{eq.curve.g2} of  potential \eqref{u.twosol} is determined by the  equation
\begin{equation*}
\nu^2=E(E-a^2)^2(E-b^2)^2.
\end{equation*}
Calculating $\nu^2(E_j)$, we have
\begin{align*}
&\nu^2(E_1)=\dfrac1{32}(a^2+b^2-B)\left((a^2-b^2)^2-B^2\right),\\
&\nu^2(E_2)=\dfrac1{32}(a^2+b^2+B)\left((a^2-b^2)^2-B^2\right).
\end{align*}
From  conditions $\nu^2(E_j)\leq 0$ ($j=1,2$) we have $B=\pm(b^2-a^2)$.

If $B=b^2-a^2$ then $E_1=b^2$, $E_2=a^2$, 
\begin{equation*}
c_2=\dfrac1{12}(c_1^2+8b^2-4a^2),\quad c_3=\dfrac1{12}(c_1^2+8a^2-4b^2),
\end{equation*}
and
\begin{align*}
&\whp_1=\whq_1=\dfrac{ib\sqrt{b^2-a^2}\cosh(at)}{(b\cosh(bt)\cosh(at)-a\sinh(bt)\sinh(at))},\quad \whq_1=-\whp_1^\ast,\\
&\whp_2=\whq_2=\dfrac{ia\sqrt{b^2-a^2}\sinh(bt)}{(b\cosh(bt)\cosh(at)-a\sinh(bt)\sinh(at))},\quad \whq_2=-\whp_2^\ast.
\end{align*}

\begin{figure}[hbtp]
\begin{center}
\begin{tabular}{p{0.45\textwidth}p{0.45\textwidth}}
\includegraphics[width=0.4\textwidth]{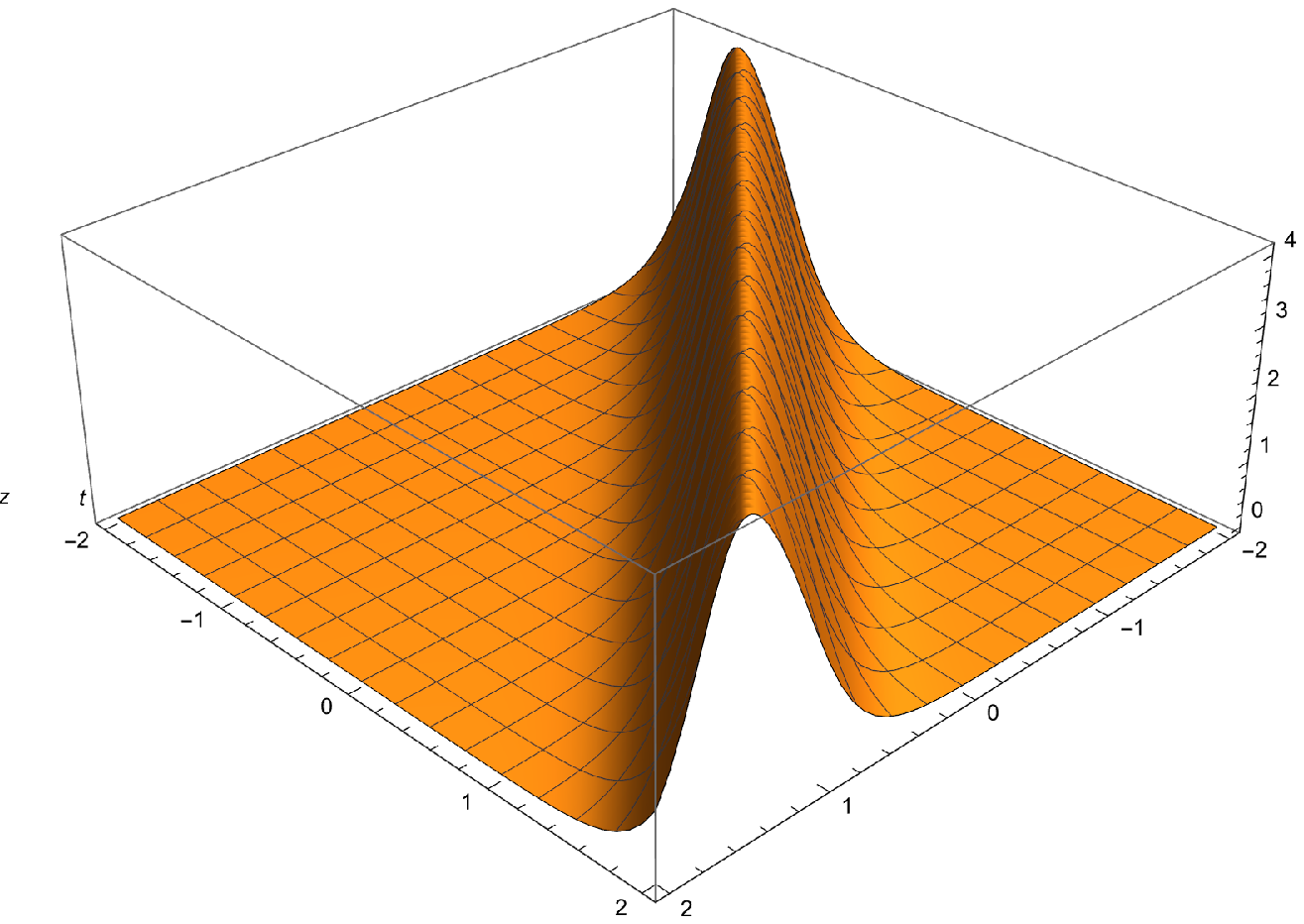}&
\includegraphics[width=0.4\textwidth]{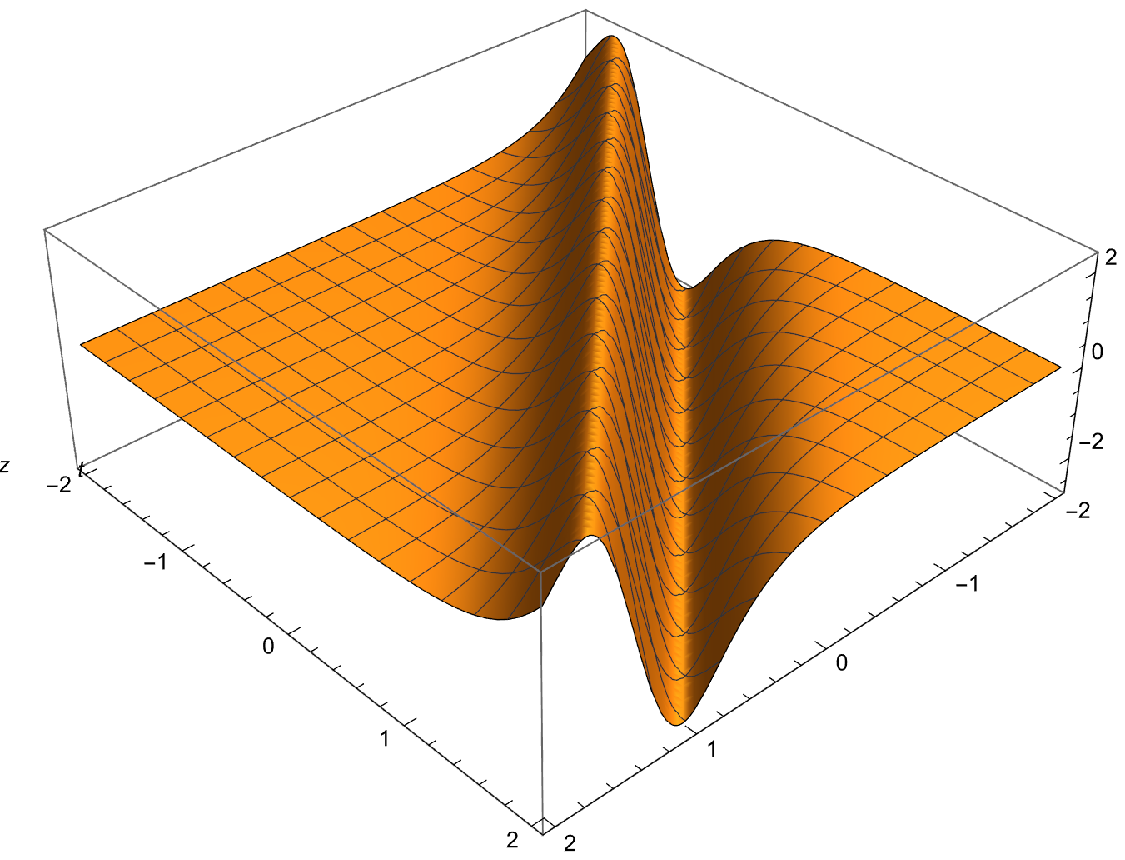}\\
\centering Magnitude $\fp_1(t-c_1z_1)$ & \centering Magnitude $\fp_2(t-c_1z_1) $
\end{tabular}
\caption{Magnitudes of  solutions  \eqref{sol.g2} for $a=3$, $b=5$, $c_1=1$, $t_a=2$, $t_b=3$.} \label{figsol.3}
\end{center}
\end{figure}

The corresponding solution to  equations \eqref{eq.man.new} has the form
\begin{equation} \label{sol.g2}
\begin{aligned}
&p_1(t,z_1)=i\fp_1(t-c_1z_1)e^{i\theta_1(t,z_1)},\quad q_1(t,z_1)=-p_1^\ast(t,z_1),\\
&p_2(t,z_1)=i\fp_2(t-c_1z_1)e^{i\theta_2(t,z_1)},\quad q_2(t,z_1)=-p_2^\ast(t,z_1),
\end{aligned}
\end{equation}
where
\begin{align*}
&\fp_1(T)=\dfrac{b\sqrt{b^2-a^2}\cosh(aT+t_a)}{(b\cosh(bT+t_b)\cosh(aT+t_a)-a\sinh(bT+t_b)\sinh(aT+t_a))},\\
&\fp_2(T)=\dfrac{a\sqrt{b^2-a^2}\sinh(bT+t_b)}{(b\cosh(bT+t_b)\cosh(aT+t_a)-a\sinh(bT+t_b)\sinh(aT+t_a))},
\end{align*}
and
\begin{align*}
&\theta_1(t,z_1)=\dfrac{c_1}2t+\left(b^2-\dfrac14c_1^2\right)z_1-\alpha\int\left(\fp_1^2(T)+\fp_2^2(T)\right)dt,\\
&\theta_2(t,z_1)=\dfrac{c_1}2t+\left(a^2-\dfrac14c_1^2\right)z_1-\alpha\int\left(\fp_1^2(T)+\fp_2^2(T)\right)dt.
\end{align*}
Here $(t_a,t_b)$ is an initial two-dimensional phase.
Magnitudes of  solutions  \eqref{sol.1} are shown in Fig.~\ref{figsol.3}.

\section{Orthogonal transformation} \label{sec.transorm}

Since the matrix $J$ has two equal diagonal elements, an orthogonal transformation of the vectors of solutions to  equations \eqref{eq.dzk} again gives solutions to these equations. To prove this statement, we consider the equation
\begin{equation} \label{lax.1T}
i\wtPsi_{t}+\wtU\wtPsi=\bo, 
\end{equation} 
where
\begin{align*}
&\wtU=\wtU_0+rJ,\quad \wtU_0=-\lambda J+\wtQ,\\
&\wtPsi=\wtT\Psi,\quad \wtT=\begin{pmatrix} 1 & \bo^t\\ \bo & T \end{pmatrix}.
\end{align*}
From  equations \eqref{lax.1} and \eqref{lax.1T} it follows that $\wtQ \wtT= \wtT Q$.
Therefore, the following equalities hold: 
\begin{equation} \label{eq.wt.pq}
\wtbq=T \bq,\quad \wtbp =\left(T^t\right)^{-1} \bp.
\end{equation}
Thus, if the matrix $T$ satisfies to the condition
\begin{equation} \label{eq.TS}
TST^\dag=S,
\end{equation}
then  vectors $\bp$ and $\bq$ ($\wtbp$ and $\wtbq$) are connected by the relation
\begin{equation*}
\bq=S\bp^\ast,\quad \wtbq=S\wtbp^\ast.
\end{equation*}

The following equalities 
\begin{equation*}
\wtG_k=T\bG_k,\quad \wtH_k=\left(T^t\right)^{-1}\bH_k,\quad \wtF_k=T F_k T^{-1},\quad \wtFf_k=\Ff_k
\end{equation*}
follow from the recurrence relations \eqref{eq.rec}.
Therefore, if  vectors $\bp$ and $\bq$ are solutions to  evolutionary equations \eqref{eq.dzk}, then  vectors $\wtbp$ and $\wtbq$ are also solutions to the same equations.
Thus, using  formula \eqref{eq.wt.pq} with the matrix
\begin{equation*}
T_1=\begin{pmatrix} \cos\varphi & \sin\varphi \\ -\sin\varphi & \cos\varphi \end{pmatrix}
\end{equation*}
 and  solution \eqref{sol.1}, 
it is possible to construct new elliptic solutions to  equation \eqref{eq.man.new}:
\begin{equation} \label{sol.wtp}
\begin{aligned}
&\wtp_1=\dfrac{i}{\sqrt{2}}\left(\cos\varphi\dn(t-c_1z_1;k)+k\sin\varphi\cn(t-c_1z_1;k)\right)e^{i\theta}, \quad \wtq_1=-\wtp_1^\ast,\\
&\wtp_2=-\dfrac{i}{\sqrt{2}}\left(\sin\varphi\dn(t-c_1z_1;k)-k\cos\varphi\cn(t-c_1z_1;k)\right)e^{i\theta}, \quad \wtq_2=-\wtp_2^\ast.
\end{aligned}
\end{equation}
%

%

Let us note that the vectors $\wtp$ and $\wtq$ satisfy the stationary equation \eqref{eq.stat} with a non-diagonal matrix 
\begin{equation*}
\wtC_n=TC_n T^{-1}.
\end{equation*}
For solution \eqref{sol.wtp} the matrix $\wtC_1$ has the form
\begin{equation*}
\wtC_1=\dfrac12\begin{pmatrix}
3(c_3+c_2)-(c_3-c_2)\cos(2\varphi)& (c_3-c_2)\sin(2\varphi)\\
(c_3-c_2)\sin(2\varphi)& 3(c_3+c_2)+(c_3-c_2)\cos(2\varphi)
\end{pmatrix}.
\end{equation*}
Therefore, for transformed solutions the constants $\wtc_j$ are equal:
\begin{align*}
&\wtc_2=\dfrac12(c_3+c_2)-\dfrac12(c_3-c_2)\cos(2\varphi),\\
&\wtc_3=\dfrac12(c_3+c_2)+\dfrac12(c_3-c_2)\cos(2\varphi),\\
&\wtc_4=\wtc_5=\dfrac12(c_3-c_2)\sin(2\varphi).
\end{align*}
At the same time, since the monodromy matrices of the functions $\wtPsi$ and $\Psi$ are similar 
\begin{equation*}
\wtM=\wtT M\wtT^{-1},
\end{equation*}
original and transformed solutions are associated  with the same spectral curve. Therefore, the Baker-Akhiezer function $\Psi$ can be constructed from a spectral curve only up to a linear transformation $\wtT$.

\section*{Discussions and conclusions}

In many works that previously investigated  finite-gap solutions of the Manakov system (see, for example \cite{EEK00, WWE07, WE07, CEEK00, EEI07, Kalla11}),  following factors that we found recently were not taken into account. First, as we showed in \cite{Sm20epj}, if  functions $p_j$ are linearly dependent, then  eigenfunctions of  Lax operator \eqref{lax.1} are determined on two separated spectral curves. Secondly, as the authors know, other researchers have not previously taken into account  preserving spectral curves orthogonal transformations of solutions. 

And finally, as we have seen from the considered in the section~\ref{sec.solutions}  examples, the number of phases of the solution is less than the genus of the corresponding spectral curve. 
Indeed, it follows from equations \eqref{eq.stat} and \eqref{eq.dzk} that the following relations hold:
\begin{equation*}
\begin{aligned}
&\partial_{z_n}\bp=-\sum_{k=1}^{n-1}c_k\partial_{z_{n-k}}\bp-c_n\partial_t\bp+i\left(r_n+\sum_{k=1}^{n-1}c_kr_{n-k}+c_n r_n\right)\bp+iC_n^t\bp,\\
&\partial_{z_n}\bq=-\sum_{k=1}^{n-1}c_k\partial_{z_{n-k}}\bq-c_n\partial_t\bp-i\left(r_n+\sum_{k=1}^{n-1}c_kr_{n-k}+c_n r_n\right)\bq-iC_n\bq.
\end{aligned}
\end{equation*}
Therefore, solutions $p_j$ and $q_j$ up to exponential multipliers are $n$-phase functions (functions with $n$ arguments):
\begin{align*}
&p_j(t,z_1,\dots,z_n)=\fp_j(t-c_nz_n,z_1-c_{n-1}z_n,\dots,z_{n-1}-c_1z_n)e^{i\theta_j(t,z_1,\dots,z_n)},\\
&q_j(t,z_1,\dots,z_n)=\fq_j(t-c_nz_n,z_1-c_{n-1}z_n,\dots,z_{n-1}-c_1z_n)e^{-i\theta_j(t,z_1,\dots,z_n)}.
\end{align*}

An equation for a spectral curve $\Gamma=\{(\mu,\lambda)\}$ has the form
\begin{equation} \label{eq.curve.mu}
\Rr(\mu,\lambda)=\det(\mu I-M)=\mu^3+\Aa(\lambda)\mu+\Bb(\lambda)=0,
\end{equation}
where
\begin{align*}
&\Aa(\lambda)=-\dfrac13\lambda^{2n+2}-\dfrac{2c_1}3\lambda^{2n+1}+\sum_{j=2}^{2n+2}A_j\lambda^{2n+2-j},\\
&\Bb(\lambda)=\dfrac{2}{27}\lambda^{3n+3}+\dfrac{2c_1}9\lambda^{3n+2}+\sum_{j=2}^{3n+3}B_j\lambda^{3n+3-j}.
\end{align*}
If $n\le 3$, then a discriminant of \eqref{eq.curve.mu}, as a polynomial of $\mu$, is equal to:
\begin{equation} \label{eq.disc}
\Delta(\lambda)=(c_{n+1}-c_{n+2})^2\lambda^{4n+4}+\sum_{j=1}^{4n+4}D_j\lambda^{4n+4-j}.
\end{equation}
Perhaps equality \eqref{eq.disc} is also true for other values of $n$. It follows from equation \eqref{eq.disc} that when condition $c_{n+1}\ne c_{n+2}$ is fulfilled, the curve $\Gamma$ has $(4n+4)$ branching points. Using the Riemann-Hurwitz formula
\begin{equation*} 
g=\dfrac{M}2-N+1,
\end{equation*}
where $M$ is a number of branching points, $N$ is a number of sheets of a covering, 
we get that the genus of the spectral curve $\Gamma$ is equal
\begin{equation*}
g=\dfrac{4n+4}2-3+1=2n.
\end{equation*}

Thus, to construct finite-gap solutions to the Manakov system or to the vector Kundu-Eckhaus equation, it is necessary to use trigonal curves, the genus of which is twice the number of phases of  solutions.

\section*{Aknowledgements} 

The research was supported by the Russian Science Foundation (grant agreement No 22-11-00196), 
\mbox{https://rscf.ru/project/22-11-00196/}



\end{document}